\documentclass{ws-ijmpd}

\begin{document}

\markboth{P. S. Negi}
{Causal configurations of homogeneous energy density}

%
\catchline{}{}{}{}{}
%

\author{P. S. NEGI} 

\title{CAUSAL CONFIGURATIONS OF HOMOGENEOUS ENERGY 
DENSITY IN GENERAL RELATIVITY}

\address{Department of Physics, Kumaun University \\
Nainital, 263 002,
India\\
negi@upso.ernet.in}

\maketitle

\begin{history}
\received{Day Month Year}
\end{history}

\begin{abstract}
\noindent
If the causality condition [the speed of sound always  remains
less  than  that  of  light  in 
vacuum, i. e., $v \leq c = 1$] is imposed on the spheres of homogeneous 
energy density,  the 
`ratio of  the  specific  heats',  $\gamma \leq  2.59457$,  constraints  the 
compaction parameter, $u [\equiv (M/a)$, mass to size ratio in geometrized 
units]  of  the  dynamically  stable  configurations $ \leq  0.34056 $
[corresponding to a surface redshift $(z_a ) \leq 0.771$].
Apparently, The maximum value of $u$ obtained in this  manner 
belongs to an absolute upper bound, and gives: (i)  The  maximum 
value for static neutron star masses as $5.4 M_\odot$, if  we  substitute 
the density at the surface  of  the  configuration  equal  to  the 
average nuclear density, $E = 2 \times 10^{14}$ g\, cm$^{-3}$ 
[e.g. {\it Nature}, {\bf 259}, 377 (1976)]. 
(ii) However,  if  the  density  of  the  static 
configuration is constrained to the value $1.072 \times 10^{14}$ g\, cm$^{-3}$,  
by imposing the empirical result that the minimum rotation period  of 
the fastest rotating pulsar known to date, PSR 1937 + 21, is 1.558 
ms, the maximum mass value for static neutron  stars  exceed  upto 
$7.4 M_\odot$. These masses have important implications for  the  massive 
compact objects like Cyg X-1, Cyg XR-1, and LMC-X3 etc., which may 
not,  necessarily,  represent  black  holes.  (iii)  The   minimum 
rotation periods for a static $1.442 M_\odot$  neutron star to  be  0.3041 
ms. (iv) A suitable stable model of ultra-compact objects [$u \ > 
(1/3)$] which has important astrophysical significance.

\end{abstract}


\section{Introduction}	

Incompressible fluid spheres of uniform energy density $E$ in 
General Relativity were first discussed by  Schwarzschild.$^1$ 
The importance of this solution in  General  Relativistic  stellar 
structures is apparent, because it gives an absolute  upper  limit 
on compaction parameter, $u (\equiv M/a$, mass to size  ratio  of  entire 
configuration in geometrized units) $ \leq (4/9) $ for any regular static 
solution in hydrostatic equilibrium.$^2$
 
     Chandrasekhar$^{3,4}$   discussed   the   condition   of 
hydrostatic equilibrium under the small  adiabatic  perturbations, 
and showed that  for  each  value  of  the  compaction  parameter, 
corresponding  to  the  {\em compressible}  homogeneous  spheres,  there 
exists a critical  (minimum)  value  of  the  ``ratio  of  specific 
heats'', $\gamma (= \gamma_{crit}$) such that for $\gamma \ <  \gamma_{crit}$,  
the  configuration 
becomes  dynamically  unstable.  For  the  limiting  case  of  the 
compaction parameter approaching  the  Schwarzschild  limit  ($u  = 
4/9), \gamma$ becomes infinity. 

     For dynamically stable superdense objects like neutron  stars 
one may expect a finite value of $\gamma$. However, for such objects  the 
equations of state are not well  known  [empirically]  beyond  the 
density $ \cong 10^{14} $ g\, cm$^{-3},^5$ and one  can  only  extrapolate 
the equations of state (available  in  the  literature)$^6$ far beyond this density. 
As  a  way  out,   one  can 
impose some restrictions upon the known physical quantities,  such 
that, the speed of sound inside the configuration, 
\begin{displaymath}
v \equiv \sqrt {(\partial {P} / \partial {E})_s}
\end{displaymath} 
(Where $P$ is the pressure, $E$ is the energy density and $s$ stands for specific
entropy) does not exceed the speed of light in vacuum, i.e., $v \leq c = 1$  (in 
geometrized units), and obtain an upper bound  on  stable  neutron 
star masses.$^{7-9}$

     In the present paper, we have  obtained  an  upper  bound  on 
compaction parameter ($u  \leq  0.34056$  corresponding  to  a  surface 
redshift  of  0.771)  for  the  compressible  homogeneous  spheres,$^{3,4}$ 
by imposing constraint on the ``ratio  of 
specific heats'', $\gamma [ \leq 2.59457)]$, compatible with causality ($v \leq 1$) 
and dynamical stability. This value of the compaction parameter is 
an absolute maximum because, for an  assigned  value  of  $\gamma$,  the 
maximum compactness would correspond  to  a  compressible  uniform 
density sphere, and can be used  to  obtain  an  upper  bound  on 
neutron star masses, as well as the minimum rotation period  of  a 
$1.442 M_\odot$  neutron star  (the  maximum  mass  of  the  neutron  star 
accurately known at present).$^{10}$

\section{ Equations Governing Radial Pulsations and Limits    
on Compaction Parameter Imposed by Causality and Ratio of the Specific Heats}

  Chandrasekhar$^{3,4}$ discussed the dynamical stability 
of  fluid  spheres  with  respect  to   small   radial   adiabatic 
oscillations on the basis  of  Einstein's  field  equations  for  a metric of the form
\begin{equation}
ds^2 =  e^{\nu} dt^2 - e^{\lambda} dr^2 - r^2 d\theta^2 - r^2 {\rm sin}^2 \theta d\phi^2 , 
\end{equation}
where $\nu$ and $\lambda$ are functions of $r$ and $t$. By  considering  the 
general time-dependent field equations appropriate for the metric 
given by Eq. (1), and letting the quantity $\xi (r)$ represent the amplitude of the 
`Lagrangian displacement' from the equilibrium position,  namely, 
$\xi(r, t) = \xi (r) e^{-i \sigma t}$, where $\sigma$  is  the  angular  frequency  of  the 
pulsation, the variational base for determining $\sigma^2$   is  given  by 
the equation$^{3,4}$
\begin{eqnarray}
\sigma^2 \int_{0}^{a} e^{(3\lambda - \nu) /2} (P + E) r^2 \xi^2 dr &  =  & 4 \int_{0}^{a} e^{(\lambda + \nu)/2} r P'\xi^2 dr  \nonumber  \\ 
                                                                   &     & + \int_{0}^{a} e^{(\lambda + 3 \nu)/2} [\gamma P/r^2] {(r^2 e^{-\nu/2} \xi)'}^2 dr  \nonumber  \\
                                                                   &     & - \int_{0}^{a} e^{(\lambda + \nu /2)} [P'^2/(P + E)] r^2 \xi^2 dr  \nonumber  \\
                                                                   &     & + 8\pi \, \int_{0}^{a} e^{(3\lambda + \nu) /2} P (P + E) r^2 \xi^2 dr.   
\end{eqnarray}

A {\it sufficient} condition for the dynamical stability of a mass is that
the right-hand side of Eq. (2) vanishes for some chosen ``trial function'' $\xi$ which
satisfies the boundary conditions
\begin{eqnarray}
\xi & = & 0 \hspace{.2in} {\rm at} \hspace{.2in} r = 0,
\end{eqnarray}
and
\begin{eqnarray}
\delta P & = & - \xi P' - \gamma P e^{\nu/2} [(r^2 e^{-\nu/2} \xi)'/r^2]  \nonumber  \\
         & = & 0 \hspace{.2in} {\rm at} \hspace{.2in} r = a,
\end{eqnarray}
where $ a $ is the size of the configuration, $\delta P$ is the `Lagrangian 
displacement  in  pressure'  and   the   prime   denotes   radial 
derivative. The quantity $\gamma$  is  defined  as$^{3,4}$
\begin{equation}
\gamma = [(P + E)/P] ({\Delta P/\Delta E})
\end{equation}
where $\Delta P$  and $\Delta E$  denote the  `Eulerian  change'  in  pressure  and 
energy density, respectively.  With  this  definition  of  $\gamma$, the 
boundary conditions [Eqs. (3) and (4)] become
\begin{eqnarray}
\xi & = & 0 \hspace{.2in} {\rm at} \hspace{.2in} r = 0,
\end{eqnarray}
and
\begin{eqnarray}
\Delta P & = & - \gamma P e^{\nu/2}[(r^2 e^{-\nu/2} \xi)'/r^2] = 0 \hspace{.2in} {\rm at} \hspace{.2in} r = a.
\end{eqnarray}
For an adiabatic perturbation, using  the  relation$^{11-13}$
\begin{equation}
\frac {\Delta P}{\Delta E} = (\partial{P}/\partial{E})_s,                                       
\end{equation}
where $s$ is the specific entropy, and if $n$ denotes the  number
density, such that, $P \equiv P(E, n)$, Eq. (5) becomes
\begin{eqnarray}
\gamma & = & (n/P) (\partial{P}/\partial{n})_s = [(P + E)/P] (\partial{P}/\partial{E})_s,                 
\end{eqnarray}
or,
\begin{eqnarray}
v^2 & = & \gamma P/(P + E) = {\rm Finite \hspace{.05in} (as \hspace{.05in} long \hspace{.05in} as} \hspace{.05in} \gamma {\rm \hspace{.05in} is \hspace{.05in} finite)}.
\end{eqnarray}
Let  us  consider   the   homogeneous   sphere   of   uniform 
energy density, $E$. The  equations  governing  equilibrium$^{3,4,14}$  can  be  written  
in  the  form  of 
compaction parameter $u$ and the radial  co-ordinate  measured  in 
units of configuration size $y (\equiv r/a)$ as 
\begin{eqnarray}
8\pi E a^2   &  =  & 6u,  \nonumber  \\
8\pi P a^2   &  =  & 6u \frac{(1-2uy^2)^{1/2} - (1-2u)^{1/2}}{3(1-2u)^{1/2} - (1-2uy^2)^{1/2}}, \nonumber \\
e^{-\lambda} &  =  & (1 - 2uy^2), \nonumber \\
e^{\nu}      &  =  & (1/4)[3(1 - 2u)^{1/2} - (1 - 2uy^2)^{1/2}]^2.
\end{eqnarray}
\newpage
\begin{table}[h]
\tbl{Various values of the compaction parameter $u$ and the corresponding critical
(minimum) values of $\gamma (\gamma_{crit})$ compatible with dynamical stability of 
{\it compressible} homogeneous sphere of uniform energy density, $E$. ${(P/E)}_0$ 
and $v_0$ represent, respectively, the pressure-density ratio and the speed of sound 
at the center of the configuration. It is seen that the dynamical stability of causal 
configurations (i. e., $u \leq 0.340555$) is constrained by the `ratio of the specific 
heats', $\gamma \leq 2.594570$.}
{\begin{tabular}{@{}ccccccc@{}} \toprule
${u}$ && $\gamma_{crit}$   &&  ${(P/E)}_0 $  && $v_0^2$  \\ \colrule
 
0.050000 && 1.38400 && 0.02780 && 0.03743 \\

0.100000 && 1.44910 && 0.06272 && 0.08552 \\

0.150000 && 1.53550 && 0.10817 && 0.14990 \\

0.200000 && 1.65620 && 0.17027 && 0.24097 \\

0.250000 && 1.83750 && 0.26120 && 0.38056 \\

0.277800 && 1.98430 && 0.33340 && 0.49615 \\

0.300000 && 2.14110 && 0.40958 && 0.62210 \\

0.320000 && 2.32890 && 0.50000 && 0.77630 \\

0.333300 && 2.49000 && 0.57710 && 0.91118 \\

0.340555 && 2.59457 && 0.62710 && 1.00000 \\ \botrule
\end{tabular}}
\end{table}

Equation (2) is evaluated for Eqs. (11) with respect  to 
the trial function, $\xi = re^{\nu/4}$ [because it gives the most  rigorous 
results$^{15,16}$  among  the  various  trial 
functions of the form $\xi = re^{\nu/N}, N = 2, 3, 4 ...\infty$,  and  the  form 
given as $\xi = b_1 r(1 + a_1 r^2  + a_2 r^4  + a_3 r^6  + ...)e^{\nu/2}$, 
where $a_1, a_2, a_3,$ ... are adjustable constants$^{17,27}$] 
for various assigned  values  of  the  constant  $\gamma$,  so  that  the 
configuration becomes compressible, and the speed of sound in this 
medium remains finite and is given by  [from  Eqs. (10)  and 
(11)]
\begin{equation}
v^2  = \gamma [1 - exp(\nu - \nu_a],
\end{equation}                                        
where the subscript $`a'$ denotes the value  of  the  corresponding 
quantity at the surface of the configuration.
 
     The   compaction    parameters    of    dynamically    stable 
configurations compatible with causality [i.e., $v \leq 1$] are given in 
Table 1. These values of $\gamma $ are consistent with those  obtained  by 
Chandrasekhar.$^{3,4}$ It is seen that the compactness of  the 
causal configuration, $u \leq 0.34056$ is constrained by the  `ratio  of 
the specific heats, $\gamma \leq 2.59457.$ Notice  that  for  a  perfectly 
incompressible homogeneous fluid sphere of uniform energy density, 
$E = n$, and the ratio of the specific heats, $\gamma$, and the speed  of 
sound in this medium become infinity for all values of $u \leq  (4/9)$, 
and the configuration would be dynamically stable for  all  values 
of $u \leq (4/9)$.

\section{ Application to Obtain an Absolute Upper Bound on Mass
and Uniform Rotation of Relativistic Stars}

By assigning the energy density  $E$  equal  to  the  average 
nuclear density, i.e. $E = 2 \times 10^{14}$ g\, cm$^{-3},^7$ in  Eqs.  (11)  for  
the  compaction  parameter,  $u = 0.34056$, we obtain a maximum mass of $5.4 M_\odot$  
[which is larger than 
the masses obtained earlier].$^{7-9}$
 
    Although,  the   assigned   value   of   energy density   thus 
substituted is reasonable, it might be  fiduciary.  Therefore,  we 
have  to  constrain  the  value  of  the  energy density  by  some 
observational fact, and it  may  be  constrained  by  the  fastest 
rotating pulsar, PSR 1937 + 20, with rotation period, $P = 1.558$ ms, 
known to  date.$^{18}$

     The  determination  of  maximum  masses  of   neutron   stars 
corresponding to maximum rotation rates require  complete  general 
relativistic calculations.$^{19,20}$  However, 
the  value  of  $E$  and  hence  the  maximum  mass  of   a   static 
(non-rotating) configuration, corresponding to higher $u$ values, can be 
obtained  very  accurately (see, e. g. Refs. 21 and 22) by 
using the empirical formula given by Koranda, Stergioulas, and  Friedman$^{23}$ 
in the following form
\begin{equation} 
P_{rot, min} {\rm (ms)} = 0.740{[M_{max}/M_\odot]}^{-1/2}{[a_{max}/10 {\rm
km}]}^{3/2}, \end{equation}
where $P_{rot, min}$ is the (minimum) rotation period corresponding to a configuration, 
rotating (uniformly) with  maximum angular velocity, and $M_{max}$ and $a_{max}$ represent,
respectively, the maximum 
mass   and   the   corresponding   size   of   the   non-rotating 
configuration. Rewriting Eq. (13)  in  terms  of  compaction 
parameter, $u[\equiv M/a]$, and angular velocity $\Omega_{max}(\equiv 2\pi/P_{rot, min})$ we 
obtain
\begin{equation}
\Omega_{max} = 2.21 \times 10^{10} [u_{max}^{1/2}/a_{max}{\rm (cm) }] {\rm s}^{-1}
\end{equation}
where $u_{max}$ is  the  maximum $u$ value  of the  non-rotating 
configuration, such that the  configuration  becomes  dynamically 
unstable  when  $u$  exceeds  $u_{max}$, and $a_{max}$    represents   the 
corresponding radius of the configuration. Note that the corresponding formula which gives an  
error of 4 - 5 \% was previously obtained by Haensel and Zdunik.$^{24}$
By the use of Eqs. (11) into Eq. (14), we  obtain
\begin{equation} 
E ({\rm g\,  cm}^{-3})  =  6.59 \times 10^6 {[\Omega_{max} ({\rm \, s}^{-1})]}^2, 
\end{equation}
and,
\begin{equation}
M_{max} {\rm (cm)}  =  [3u_{max}^3/4\pi E ({\rm cm}^{-2})]^{1/2}.
\end{equation}

Thus, the {\it uniform} energy density of the configuration depends only  upon 
the rotation period, and not upon the compaction  parameter $ u $. 
Therefore, it is clear from Eqs. (15) and (16)  that  for  a 
given value of the rotation period $P_{rot}$ the maximum mass of the 
stable configuration depends only upon the maximum value of $ u (u_{max})$. 
For $P_{rot} = 1.558$ ms, Eq. (15)  gives  the  energy density, 
$E$,  of  the  configuration  as $1.072  \times 10^{14}$ g\, cm$^{-3}\,$,   
the substitution of $u_{max} \cong  0.34056 \,$  from  Table 1 into Eq. (16)
gives the   maximum mass of the configuration, $M_{max} \cong 7.387 M_\odot$ 
and the corresponding  radius, $a_{max} \cong 31.974$ km. Notice that the
range of energy density obtained in this   manner is applicable to the
baryonic equation of state, known as  $Q$-Star  equation of state.$^{5,22,25}$

For the value of $u \cong 0.34056$, the speed of sound is maximum,  $v \cong 1$,
at the center and decreases monotonically  from  center  to  the
surface of the configuration, and at the surface it vanishes along 
with  the  pressure.  These  masses  have  important  implications 
regarding massive compact objects like Cyg X-1, Cyg XR-1,  and  LMC  X-3 etc. 
These limits of maximum masses could  be  updated  if  the  faster 
rotating pulsars are observed in the future [see, Ref. 22 for a
detailed discussion in this regard]. 

     On  the  other  hand,  if  we  impose   the   constraint   on 
energy density [by using Eq. (16)] such that the maximum mass of neutron star is $1.442 
M_\odot,^{10}$ then Eq. (15) gives the  minimum 
rotation period 0.3041 ms for the maximum $ u  \cong  0.34056$,  and  the 
corresponding uniform energy density $E$ is obtained  as $ 2.813  \times 
10^{15} $  g\, cm$^{-3}$.

\section{ Results and Conclusions}
     An absolute upper bound on compaction parameter, $u  \leq  0.34056 $
[or the surface redshift $ \leq 0.771$], compatible with  causality  and 
the ratio of the specific heats, $\gamma \leq 2.59457$, is obtained by  using 
the dynamically stable compressible homogeneous sphere. This upper 
limit of compaction parameter gives

(i) The maximum static mass of conventional model of neutron stars 
[taking $E = 2 \times 10^{14}$ g\, cm$^{-3}]^{7,8}$ as $5.4 M_\odot$. This is greater than $4.8 M_\odot$   considered  as  an 
upper limit earlier.

(ii) The maximum mass of static neutron star exceeds to the  value 
of $7.387 M_\odot$  which is greater than the upper limit of $5.3 M_\odot$ for 
neutron stars (so called $Q$-star models) obtained by Hochron,  Lynn 
and Selipesky.$^{26}$ This may have important implications for  the 
heavy compact objects like Cyg X-1, Cyg XR-1, and LMC-X3 which may 
not, necessarily, be black holes.
 
(iii) The minimum rotation periods for a static $1.442 M_\odot$   neutron 
star to be 0.3041 ms with a uniform energy density $E$ as $2.813  \times 
10^{15}$ g\, cm$^{-3}$.

(iv) A causally consistent and dynamically stable model  of  ultra-compact objects [$u \ >  (1/3)$]  which  has  important  astrophysical 
significance [see, e.g. Ref. 16 and  references therein].

\section*{Acknowledgments}
     The author acknowledges State Observatory, Nainital for providing library and computer center facilities.

\end{document}